\documentclass[pra,twocolumn,amsmath,amssymb,groupedaddress,superscriptaddress]{revtex4}

\usepackage{graphics,xcolor}
\usepackage{amsmath}
\usepackage{bm}
\usepackage{float,graphicx}
\usepackage{subfigure}
\usepackage{epstopdf}
\usepackage{soul}
\usepackage{dsfont}
\usepackage{booktabs}
\usepackage[normalem]{ulem}
\usepackage{footnote}
\usepackage{mathtools}
\usepackage{tikz}
\usepackage{braket}
\usepackage{xfrac}
\usepackage{multirow}
\usepackage[linkcolor=black, urlcolor=black,colorlinks=true, citecolor=cyan]{hyperref}

\newcommand{\wrt}{w.\,r.\,t.~}
\newcommand{\eq}{eq.~}
\newcommand{\aop}{a}
\newcommand{\cop}{a^{\dagger}}

\newcommand{\Ord}{O\!\left( |\dtk|^2 \right)}
\renewcommand{\O}[1]{\mathcal{O}\big(#1\big)}

\newcommand{\Htfi}{\mathcal{H}_\mathrm{TFI}}

\newcommand{\Hjj}{\mathcal{H}_\mathrm{J1J2}}

\newcommand{\bx}{\bm{x}}

\newcommand{\pars}{\theta}
\newcommand{\wf}{\psi_{\pars}}
\newcommand{\wfnqs}{\wf}
\newcommand{\wfi}{\psi_{\epsilon}}
\newcommand{\wfd}[1]{\psi_{#1,\pars}}
\newcommand{\wfe}{\psi_{\pars + \Delta \pars}}

\newcommand{\nwf}[1]{\overline{\psi}_{#1,\pars}}
\newcommand{\nwfd}[1]{\overline{\psi}_{#1,\pars}}
\newcommand{\nwfdt}[1]{\widetilde{\psi}_{#1,\pars}}

\newcommand{\nwfe}{\overline{\psi}_{\pars + \Delta \pars}}
\newcommand{\nwfet}{\widetilde{\psi}_{\pars + \Delta \pars}}

\newcommand{\Nv}{n_\mathrm{var}}
\newcommand{\Ns}{n_\mathrm{s}}

\newcommand{\wfrbm}{\psi_{\bm{\lambda}}^{\mathrm{rbm}}}

\newcommand{\Nvmc}{N_{\mathrm{vmc}}}

\newcommand{\evmc}{E_\mathrm{vmc}}

\newcommand{\tu}{t_\mathrm{u}}
\newcommand{\ts}{t_\mathrm{s}}

\newcommand{\ttrans}{t_s^\mathrm{trans}}

\newcommand{\convr}{c_r}

\newcommand{\Dd}[1]{\mathcal{D}_{#1}^\dagger}
\newcommand{\D}[1]{\mathcal{D}_{#1}}
\newcommand{\Hl}{\mathcal{H}_\mathrm{loc}}
\newcommand{\Hld}[1]{\mathcal{H}_{\mathrm{loc}, #1}}

\newcommand{\sexpb}[1]{\mathbb{E} \big [ #1 \big]_\mathcal{M}}

\newcommand{\dt}{\Delta \pars}
\newcommand{\dtk}{\Delta \pars_k}

\newcommand{\pd}{\partial_{\pars_k}}
\newcommand{\pdd}{\partial_{\pars_k^*}}

\newcommand{\ti}{t_i}
\newcommand{\Tt}{T}

\newcommand{\Tu}{T_u}

\newcommand{\nconv}{n_\mathrm{conv}}
\newcommand{\nis}{\nconv}

\newcommand{\erel}{\epsilon_\mathrm{rel}}

\begin{document}
\title{Learning Neural Network Quantum States with the Linear Method}
\author{J. Thorben Frank\footnote{thorbenjan.frank@gmail.com}}
\affiliation{Institute for Theoretical Physics, University of Cologne, Zülpicher Str. 77, 50937 Cologne, Germany}
\affiliation{Machine Learning Group, Technical University of Berlin, Marchstr. 23, 10587 Berlin, Germany}
\author{Michael J. Kastoryano}
\affiliation{Institute for Theoretical Physics, University of Cologne, Zülpicher Str. 77, 50937 Cologne, Germany}
\affiliation{Amazon Quantum Solutions Lab, Seattle, Washington 98170, USA}
\affiliation{AWS Center for Quantum Computing, Pasadena, California 91125, USA}
\begin{abstract}
Due to the strong correlations present in quantum systems, classical machine learning algorithms like  stochastic gradient descent  are often insufficient for the training of neural network quantum states (NQSs). These difficulties can be overcome by using physically inspired learning algorithm, the most prominent of which is the stochastic reconfiguration (SR) which mimics imaginary time evolution. Here we explore an alternative algorithms for the optimization of complex valued NQSs based on the linear method (LM), and present the explicit formulation in terms of complex valued parameters. Beyond the theoretical formulation, we present numerical evidence that the LM can be used successfully for the optimization of complex valued NQSs, to our knowledge for the first time. We compare the LM to the state-of-the-art SR algorithm and find that the LM requires up to an order of magnitude fewer iterations for convergence, albeit at a higher cost per epoch. We further demonstrate that the LM becomes the more efficient training algorithm whenever the cost of sampling is high. This advantage, however, comes at the price of a larger variance. 
\end{abstract}
\maketitle
\section{Introduction}
Neural network quantum states (NQSs) \cite{carleo2017solving} constitute one of the most promising approaches for the description of quantum many-body systems. Since their introduction in 2017, they have been successfully applied to various lattice systems in one and two dimensions \cite{choo2018symmetries,choo2019two, ferrari2019neural, yang2020scalable}. More recently they have also been applied to fermionic systems in second \cite{choo2019fermionic} and first quantization \cite{stokes2020phases}. 

A variety of  different network architectures has been explored for use as a variational Ansatz class; including Restricted Boltzmann Machines (RBMs) \cite{carleo2017solving, smolensky1986information},  convolutional neural networks  \cite{lecun1998gradient, choo2019two, schmitt2020quantum}, and more recently recurrent neural networks \cite{hibat2020recurrent}, deep auto-regressive networks \cite{sharir2020deep} and graph neural networks  \cite{sperduti1997supervised, yang2020scalable}.

Although there is a diverse array of NQS architectures available, essentially all of the high precision ground state projection algorithms employ the Stochastic Reconfiguration (SR) \cite{sorella2001generalized} learning method. During the optimization, the SR updates the Ansatz wave function to mimic imaginary time evolution within the variational subspace, hence projecting onto the ground state. This corresponds to a stepwise rescaling of  the gradient according to the quantum fisher matrix \cite{park2020geometry}. Thus, the learning is closely related to the natural gradient descent method \cite{amari1998natural}.

\begin{figure}
    \centering
    \includegraphics[width=\linewidth]{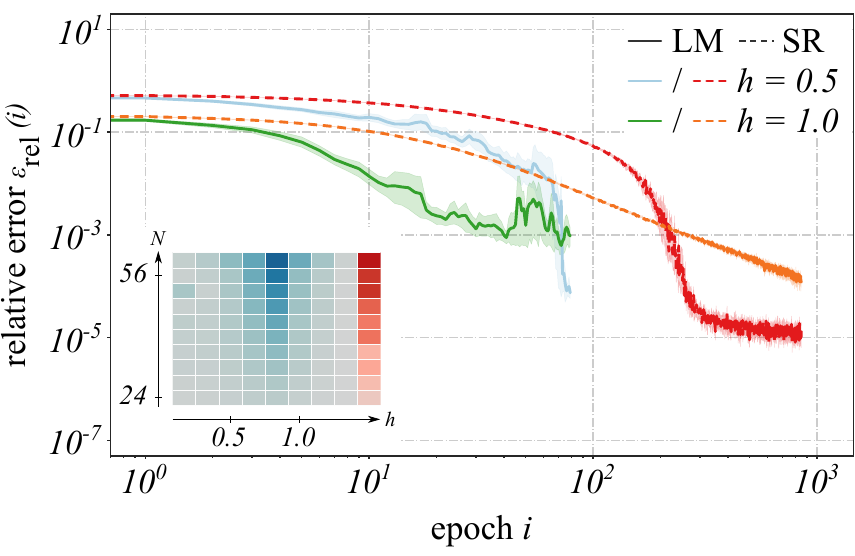}
    \caption{\textbf{Learning curves with LM and SR: } Relative error $\erel(i)$ as a function of the VMC iteration $i$, where  $\erel(i) = |E_0 - \evmc(i)|/|E_0|$ is the deviation of the VMC estimate from the exact ground state energy for the transverse field Ising model in one dimension. Plotted are the learning curves for training a symmetric RBM ($\alpha = 2$) with the Linear Method (LM) (for 75 epochs) and the Stochastic Reconfiguration (SR) (for 750 epochs) for different strengths of the external field $h$ for a chain with $N = 24$ sites.  The inset in the figure shows the relative computational cost of optimization as a function of external field and chain length. Red indicates settings where the SR is more efficient, and blue indicates when the LM is more efficient. The color scale is linear.}
    \label{fig:vmc-curves-accuracy-lattice}
\end{figure}

There are two essential aspects in the evaluation of neural network architectures: (1) \textit{expressivity} of the Ansatz and (2) \textit{efficiency} of the learning scheme \cite{livni2014computational}. While a lot of effort has been put into the characterization of expressivity for different NQS architectures \cite{glasser2018neural,levine2019quantum, sehayek2019learnability}, the efficiency question has not received nearly as much attention. At present, systematic improvements to the training time can only be achieved by adjusting SR hyperparameters or by reducing the number of model parameters. The latter, however, results in a decrease of the expressivity of the NQS. The possibility of changing the training algorithm itself is currently inaccessible since at present, SR is the only reliable learning algorithm for NQSs. We take this as a motivation to implement a  tailored learning algorithm for the optimization of NQSs, which is based on the Linear Method (LM), first developed in the context of Variational Monte Carlo (VMC) \cite{umrigar2005energy,umrigar2007alleviation}. While, VMC typically uses real variational parameters our results are, to the best of our knowledge, the first investigating the LM for wave functions with complex valued parameters.

In this paper we investigate the scaling of the computational cost when training NQSs with the LM and compare it with the SR. Such scaling analysis is ambiguous in the VMC setting which usually considers molecular systems, where it is difficult to distinguish effects arising form increasing the size of the simulation from those originating in the properties of the specific molecular orbital basis choice.

We find that the LM typically requires up to an order of magnitude fewer training epochs for learning the ground state, albeit at a higher cost per epoch. Thus, one faces a typical trade-off problem when searching for the more efficient learning algorithm. Indeed we find this problem to be non-trivial since it depends on the Hamiltonian under investigation as well as on the point in the Hamiltonian parameter space.


Our scaling analysis allows for the identification of a crossover point in which the choice for the more efficient training algorithm changes. Starting from there it can be shown that the location of this crossover point strongly depends on the computational cost of sampling. In this context sampling corresponds to the generation of samples $\{\bx_k\}$ \wrt the absolute square of the parametrized wave function $|\wf(\bx)|^2$ using Markov chain Monte Carlo (MCMC). More specifically we find that as long as sampling is easy (computationally cheap) the SR is more efficient. When sampling is expensive, however, the LM shows clear advantages with respect to computational training cost. These qualitative statements are substantiated in the main body of text.

Thus, we expect the LM to be a valuable tool for the training of NQSs whenever sampling requires large amounts of computational resources. We will illustrate the strength of the LM on the example of fermionic NQSs for simple chemistry Hamiltonians. There it has been shown that a tremendous amount of samples is necessary to generate sufficient statistics \cite{choo2019fermionic}, which is due to the sharply peaked nature of the wave function around the Hartree-Fock state.
Beyond that, our results are numerical evidence that the LM can be used for the learning of complex valued NQSs expanding the set of  reliable and accurate learning algorithms. 


\section{Theory}
\subsection{Neural Network Quantum States} \label{sec:nqs}
The basic idea of NQSs relies on parametrizing some quantum state vector in a basis $\{\bx\}$ as
\begin{align}
	\ket{\wfnqs} = \sum_{\bx} \wfnqs(\bx) \ket{\bx},
\end{align}
such that the complex amplitude of the wave function $\wf(\bx)$ is defined by the output of some neural network. 

The parameter configuration $\pars^*$ describing the ground state can be written as a minimization problem 
\begin{align}
	E_{\pars^*} = \underset{\pars}{\mathrm{min}} \, \braket{\mathcal{H}}_{\pars}, \label{eq:energy-minimization}
\end{align}
where $E_{\pars^*}$ is the global minimum of the variational energy $\braket{\mathcal{H}}_{\pars} \equiv \braket{\wf|\mathcal{H}|\wf}$ in the $\Nv$-dimensional space.

As the exact evaluation of  expectation values, and thus of the variational energy, scales exponentially in the system size, they are estimated by MCMC sampling 
\begin{align}
	\braket{\mathcal{O}} \simeq \sexpb{\mathcal{O}}, \label{eq:stochastic-estimation}
\end{align}
for some (local) observable $\mathcal{O}$, where $\sexpb{\mathcal{O}} = \sum_{\bx \in \mathcal{M}} \mathcal{O}(\bx) |\wfnqs(\bx)|^2$ is the classical expectation value for the current wave function $\wf(\bx)$. Thus, the evaluation of $\sexpb{\cdots}$ requires the generation of samples $\mathcal{M} = \{\bx\}$ distributed \wrt the absolute square of the current parametrized wave function $|\wfnqs(\bx)|^2$.

The value of the variational energy for some given parameter configuration can be calculated as 
\begin{align}
	\braket{\mathcal{H}}_\pars \simeq \sexpb{\Hl},
\end{align}
where $\Hl(\bx)$ is the so called \textit{local energy}
\begin{align}
	\Hl(\bx) = \frac{\braket{\bx|\mathcal{H}|\wfnqs}}{\braket{\bx|\wfnqs}}.
\end{align}
Note that the local energy in principle depends on the point in the parameter space but its dependence on $\pars$ has been left out for notational convinience. 

The global minimum of equation \eqref{eq:energy-minimization} is found by iteratively updating the variational parameters $\pars$
\begin{align}
\pars^{(i+1)} \leftarrow \pars^{(i)} + \dt^{(i)},
\end{align}
where the update step $\dt$ is calculated based on local approximations made to the variational wave function $\wf$ and energy $\braket{\mathcal{H}}_\pars$. In the following we will review the two main learning algorithms that will be analyzed throughout this paper, the Stochastic Reconfiguration (SR) and the Linear Method (LM). 
Given that the LM is usually applied to real valued wave functions one has to carefully consider the complex case in order to make the LM applicable to NQSs. 
\subsection{Stochastic Reconfiguration}
Originally introduced by Sorella in 2001 \cite{sorella2001generalized} for the optimization of Jastrow wave functions, the SR is currently the state-of-the-art method for the optimization of NQSs for quantum spin systems. The idea is to update the parameters in the variational wave function, such that it mimics the imaginary time evolution $ \ket{\psi_{\tilde{\theta}}} \approx e^{-\epsilon \mathcal{H}}\ket{\wf}$ \cite{glasser2018neural}. Beside its physical foundation, it is also closely related (and for positive wave functions equivalent) to the natural gradient descent \cite{amari1998natural}, a learning algorithm originated in the machine learning community. 

Within the SR each update step is calculated as  
\begin{align}
	\pars' \leftarrow \pars - \eta \bm{S}^{-1} \bm{f}, \label{eq:sr-update}
\end{align}
where $\eta$ is the so-called learning rate, $\bm{S}$ the quantum Fisher matrix
\begin{align}
	S_{kk'} = \braket{\Dd{k} \D{k'}} - \braket{\Dd{k}} \braket{\D{k'}}, \label{eq:s-matrix-sr}
\end{align}
and $\bm{f}$ the force vector
\begin{align}
	f_k = \braket{\Dd{k} \mathcal{\Hl}} - \braket{\Dd{k}}\braket{\mathcal{\Hl}}. \label{eq:force-vector-sr}
\end{align}
For the readers convenience, 
a derivation of Eqns. (\ref{eq:s-matrix-sr},\ref{eq:force-vector-sr}) and of the the stochastic estimators is provided in appendices \ref{app:sr-as-imaginary-time-evolution} and \ref{app:stochastic-estimation-of-quantum-expectation-values}. The quantity $\D{k}$ in the equations above describes the change in the wave function with the variational parameters such that $\pd \ket{\wfnqs} = \D{k} \ket{\wfnqs}$, where $\pd$ is the partial derivative \wrt to the $k$-th variational parameter $\pars_k$. Its corresponding stochastic estimator is the so-called \textit{log-derivative} which is defined as 
\begin{align}
	\D{k}(\bx) = \pd \log \big[ \wfnqs(\bx) \big] \label{eq:log-derivatives}.
\end{align}

The $\bm{S}$ matrix can be non-invertible due to vanishing eigenvalues. In order to ensure invertibility of $\bm{S}$, a positive regularization constant $a_\mathrm{diag}$ is added to its diagonal as
\begin{align}
\widetilde{\bm{S}} = \bm{S} + a_\mathrm{diag} \bm{1}, \label{eq:sr-diag-shift}
\end{align}
where $\bm{1}$ is the identity.
\subsection{Linear Method}
The LM was originally introduced by Nightingale and Melik-Alaverdian \cite{nightingale2001optimization} in 2001 and was later extended by Umrigar et al \cite{umrigar2007alleviation} to allow for the optimization of non-linear parameters. At its core, the LM relies on the linear expansion of the explicitly normalized wave function, which lies in the self-plus tangent space $\Omega^+$ \cite{zhao2017blocked}, spanned by the current normalized wave function $\ket{\nwfd{0}}$ and its tangent space $\Omega \coloneqq \big\{ \ket{\psi} \mid \braket{\psi | \nwfd{0}} = 0\big\}$. Solving the projected Schrödinger equation in $\Omega^+$ results in update steps which often outperform second order methods like the Newton-Raphson method, as they obey a strong zero variance property.

Lets start by considering the first order expansion of the explicitly normalized wave function $\ket{\nwfd{0}}$ which can be written as
\begin{align}
\ket{\nwfe} = \exp\big[  i \Phi \big]\sum_{k = 0}^{\Nv} \dtk \ket{\nwfd{k}}, \label{eq:lin-exp-wf}
\end{align}
where $\Phi$ is a global phase factor and $\ket{\nwfd{0}}$ is the current normalized wave function with $\dt_0 = 1$. For $k > 0$ the derivative \wrt the $k$-th variational parameter is given as
\begin{align}
\ket{\nwfd{k}} = \big[ \D{k} - \braket{\D{k}} \big]\ket{\nwfd{0}}, \label{eq:tangent-space-derivative}
\end{align}
where $\D{k}$ are the log-derivatives \eqref{eq:log-derivatives}. It can then be easily verified that the $\ket{\nwfd{k}}$ lie in the tangent space $\Omega$ of $\ket{\nwfd{0}}$ for $k > 0$ such that $\ket{\nwfe}$ lies in the self-plus tangent space $\Omega^+$. While the orthogonality between $\ket{\nwfd{0}}$ and $\ket{\nwfd{k}}$ is straightforward to show in the case of real parameters, the case of complex parameters generally results in a non-zero overlap between the current wave function and its derivatives, as pointed out by Motta et al \cite{motta2015implementation}. However, up to first order, orthogonality can be restored for the complex case using phase optimization (see appendix \ref{app:wave-function-phase-optimization}). 

With the linear expansion at hand, the idea of the LM relies on minimizing the linear energy approximation
\begin{align}
E_{\mathrm{lin}}^* &= \underset{\dt}{\mathrm{min}}\,\, E_\mathrm{lin}(\dt)\\ &= \underset{\dt}{\mathrm{min}}\,\, \frac{\braket{\nwfe|\mathcal{H}|\nwfe}}{\braket{\nwfe|\nwfe}} \label{eq:lm-minimization},
\end{align}
where we define $\overline{\bm{H}} \equiv \braket{\nwfe|\mathcal{H}|\nwfe}$ and $\overline{\bm{S}} \equiv \braket{\nwfe|\nwfe}$ which are the so-called Hamilton and overlap matrix, respectively. The overlap matrix is fully expressed in terms of the log-derivatives $\D{k}$
\begin{align}
\overline{S}_{00} &= 1, \label{eq:lm-s00}\\
\overline{S}_{k0} &= \overline{S}_{0k'} = 0, \label{eq:lm-sk0-s0k}\\
\overline{S}_{kk'} &= \braket{\Dd{k} \D{k'}} - \braket{\Dd{k}} \braket{\D{k'}}, \label{eq:lm-quantum-fisher}
\end{align}
where one identifies the quantum Fisher matrix \eqref{eq:s-matrix-sr} in the last expression. The entries of the Hamilton matrix also include the log-derivatives as well as additionally the Hamiltonian operator
\begin{align}
H_{00} &= \braket{\mathcal{H}}, \label{eq:lm-h00}\\
H_{k0} &= \braket{\Dd{k} \mathcal{H}} - \braket{\Dd{k}}\braket{\mathcal{H}}, \label{eq:lm-hk0}\\
H_{0k'} &= \braket{\mathcal{H} \D{k'}} - \braket{\mathcal{H}} \braket{\D{k'}}, \label{eq:lm-h0k}\\
H_{kk'} &= \braket{\Dd{k} \mathcal{H} \D{k'}}  - \braket{\Dd{k} \mathcal{H}} \braket{\D{k'}} \nonumber\\ &\hspace{2em}- \braket{\Dd{k}} \braket{\mathcal{H} \D{k'}} + \braket{\Dd{k}} \braket{\mathcal{H}} \braket{\D{k'}}. \label{eq:lm-hkk}
\end{align}
The  expectation values can be estimated stochastically as in Eqn. \eqref{eq:stochastic-estimation} which is shown in detail for equations \eqref{eq:lm-s00} - \eqref{eq:lm-hkk} in appendix \ref{app:stochastic-estimation-of-quantum-expectation-values}.

The energy minimization as expressed by equation \eqref{eq:lm-minimization} is a quadratic problem in the parameter change $\dt$, where it will always hold that $\braket{\nwfe|\nwfe} \geq 1$ due to $\ket{\nwfd{k}} \in \Omega$ for $k > 0$. Thus, equation \eqref{eq:lm-minimization} expresses the trade-off between energy minimization (making $\braket{\nwfe|\mathcal{H}|\nwfe}$ as small as possible) and normalization conservation (keeping $\braket{\nwfe|\nwfe}$ as close to one as possible). The solution to \eqref{eq:lm-minimization} can be found as the vector $v_0 \in \mathbb{C}^{|\pars|}$ associated with the smallest real eigenvalue $\lambda_0$ of the generalized eigenvalue equation 
\begin{align}
\overline{\bm{H}}\, \begin{bmatrix}
c \\
v_0
\end{bmatrix} = \lambda_0 \, \, \overline{\bm{S}} \, \begin{bmatrix}
c \\
v_0
\end{bmatrix}, \label{eq:lm-gen-eigenvalue-prob}
\end{align}
where $c$ is a complex scalar. In order to ensure invertibility of $\overline{\bm{S}}$ in a similar fashion as in equation \eqref{eq:sr-diag-shift} a shift is added to the diagonal of $\overline{\bm{S}}$ apart from the first entry. The update within the LM is then given as
\begin{align}
\pars' \leftarrow \pars + v_0/c. \label{eq:lm-update}
\end{align}

\subsubsection{Stabilization by Regularization}\label{sec:tikhonov-reg} 
Let $\ket{\psi^{'}}$ be the new wave function that has been updated according to equation \eqref{eq:lm-update}. Then its normalized form will be a good approximation of the subspace eigenfunction $\ket{\nwfe} \in \Omega^+$, as long as the update step taken in the variational space is small enough \cite{zhao2017blocked}. Sufficiently small update steps can be ensured by adding a positive, real constant $\kappa$ to the diagonal of the $\overline{\bm{H}}$ matrix (except for the first entry) \cite{toulouse2007optimization}. This regularization scheme plays the same role as the trust radius in the Newton-Raphson optimization and can be interpreted as damping of the update step, which means increasing $\kappa$ reduces the length of $\delta \bm{\lambda}$. The optimal choice of $\kappa$ is by far not obvious and depends on the current point in the optimization as well as the system under investigation as it determines the energy landscape. A common way is to solve the eigenvalue problem defined in equation \eqref{eq:lm-gen-eigenvalue-prob} for three different values of $\kappa$
\begin{align}
    \widetilde{H}_{k,k'} = H_{k,k'} + \kappa_n \, \delta_{k,k'} (1 - \delta_{k,0}), \label{eq:tikhonov-reg}
\end{align}
with $\kappa_n = \kappa_0 \times 10^{n}$. Afterwards, one chooses the update step that yields the lowest variational energy, which can be estimated over the correlated sampling technique described in \cite{filippi2000correlated}.
\subsection{Algorithmic Outline}
For the SR and the LM, each training iteration consists of the following steps:
\begin{enumerate}
    \item[(1)] Generate $\Ns$ samples with respect to $|\wfnqs(\bx)|^2$ using the MCMC sampling.
    \item[(2)] Calculate the stochastic estimates $\sexpb{\cdots}$ for the quantum expectation values $\braket{\cdots}$ that appear in the update matrices.
    \item[(3)] Construct the components of:
    \begin{itemize}
        \item[-] The overlap matrix $\overline{\bm{S}}$ and Hamilton matrix $\bm{H}$ (LM).
        \item[-] The quantum Fisher matrix $\bm{S}$ and force vector $\bm{f}$ (SR).
    \end{itemize}
    \item[] Solve the  linear set of equations for $\dt$.
    \item[] Update the network parameters as: $\pars' \leftarrow \pars + \dt$.
\end{enumerate}
Steps (1) - (3) are repeated until the energy $E_\pars$ is considered optimal. A detailed analysis of the computational complexity can be found in the appendix \ref{app:computational-complexities}.
\subsection{RBM Quantum States}
Throughout, we will consider RBM quantum states, parametrized as $\ket{\wf} = \sum_x \wf(x) \ket{x}$ where the complex amplitude associated with the basis states is given by the exponential family 
\begin{align}
    \wf(x) = \sum_{y \in \{0,1\}^M} \exp \left(x^{T} \cdot w \cdot y + a \cdot x + b \cdot y\right). \label{eq:spin-rbm-function}
\end{align}
The binary vectors $x$ and $y$ of length $N$ and $M$ represent the visible and the hidden state, respectively. In the context of RBM quantum states the visible state is taken to be the basis state $x$. Each connection between hidden $y_i$ and visible node $x_j$ has an associated weight $w_{ij}$ and the vectors $a \in \mathbb{C}^N$ and $b \in \mathbb{C}^M$ describe the biases attached to visible and hidden nodes, respectively.
\begin{figure}
    \centering
    \includegraphics[width=\linewidth]{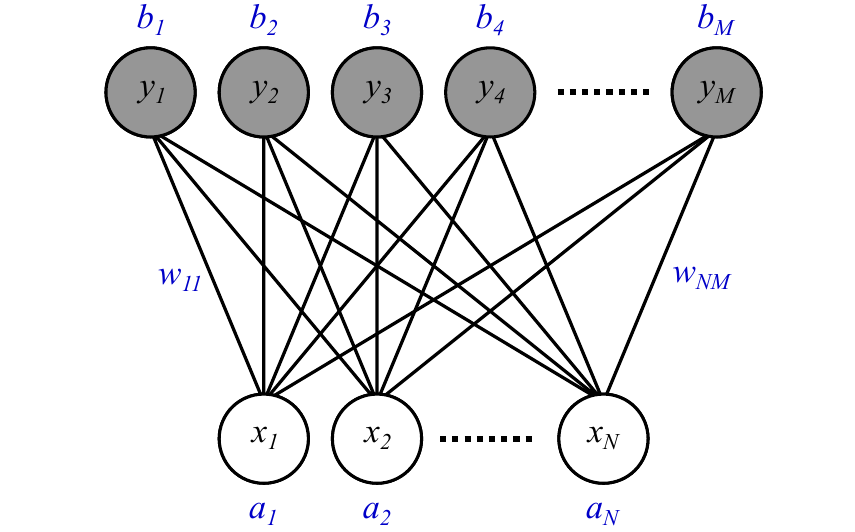}
    \caption{An RBM consisting of a visible and a hidden unit layer, where states are given as the binary vectors $x$ and $y$ of length $N$ and $M$, respectively. The bias vectors for the hidden and visible unites are given as $b \in \mathbb{C}^M$ and $a \in \mathbb{C}^N$. The weights on the connections between visible and hidden units are the entries of the matrix $w \in \mathbb{C}^{N\times M}$.}
    \label{fig:rbm}
\end{figure}
We denote the whole set of variational parameters in the RBM quantum state as $\pars = \{a,b,\mathrm{vec}(w)\}$ with weight matrix $w \in \mathbb{C}^{N\times M}$. The ratio between number of hidden and visible units $\alpha = M/N$ is commonly referred to as hidden unit density and enters the total number of variational parameters via $\Nv = \alpha N^2 + \alpha N$ scaling quadratically in the system size $N$. For systems exhibiting full translational symmetry
the effective number of variational parameters can be reduced by one order of magnitude to $\Nv \propto \mathcal{O}(\alpha N)$ (for more details see \cite{carleo2017solving}).
The optimization of RBM quantum states with the SR and the LM is computationally efficient when the following requirements hold \cite{park2020geometry}:
\begin{enumerate}
	\item[(1)] The computation of the operators $\D{k}(\bx)$, $\Hl(\bx)$ (and $\Hld{k}(\bx)$ for the LM) is efficient for every state $\bx$.
	\item[(2)] Efficient sampling with respect to $\wf(\bx)$ is possible for any values of $\theta$. This means the updates in a Monte Carlo chain can be computed efficiently, as the ratio of wave functions $\wf(\bx)$ and $\wf(\bx')$ can be computed efficiently for any $\bx$ and $\bx'$. 
	\item[(3)] The Markov chain converges to the desired state in sub-polynomial time.
\end{enumerate}
Although requirement (3) can hardly ever be guaranteed one observes in practice that it is often satisfied, as long as the Hamiltonian is free of frustration (note however recent proposals involving auto-regressive learning strategies, that guarantee efficient sampling \cite{sharir2020deep}). Indeed it should be noted, that the requirements above are sufficient for the learning algorithms presented in this paper, but might be altered if one is using other, e.g. second order methods, that additionally rely on the efficient calculation of second order derivatives $\D{k,k'}(\bx)$.

\subsection{Spin Models}

Throughout this paper, the LM and SR  will be compared for the transverse field Ising (TFI) model as well as for the J1J2 model. In this context $X,Y,Z$ will denote the Pauli operators with corresponding eigenvalues $x,y,z = \pm 1$ and periodic boundary conditions are assumed for all models. Beyond that, we apply local basis transformations to make the TFI model stoquastic and the J1J2 model less frustrated. This has been shown to improve the performance when training NQSs \cite{park2020neural}.

\subsubsection{Transverse Field Ising Model}
The Hamiltonian of the TFI model is given as
\begin{align}
\Htfi = - \sum_{\braket{i,j}} Z_i Z_j - h \sum_i X_i \label{eq:ham-tfi}
\end{align}
where $\braket{\cdot\,,\cdot}$ denotes nearest neighbours and $h$ the strength of an external field, which is applied perpendicular to the $z$ axis. At $h = 1$ the TFI model has a quantum critical point while it reduces to the classical Ising model for $h = 0$. In 1D, the model has an exact solution by mapping it to free Fermions \cite{pfeuty1970one}.

\subsubsection{J1J2 Model}
The second spin model that will be investigated is the J1J2 model which includes next nearest neighbour interactions. Its Hamiltonian is parametrized by the nearest and the next nearest neighbour couplings $J_1$ and $J_2$:
\begin{align}
\Hjj &= J_1 \sum_{\braket{i,j}} X_i X_j + Y_i Y_j + Z_i Z_j \nonumber\\&\hspace{2.5em}+ J_2 \sum_{\braket{\braket{i,j}}} X_i X_j + Y_i Y_j + Z_i Z_j,
\end{align}
where $\braket{\braket{\cdot\,,\cdot}}$ indicates next nearest neighbours. Throughout the paper, we fix $J_2 = 1$ and restrict $J_2 \geq 0$ resulting in anti-ferromagnetic interactions such that the total magnetization of the ground state is zero, allowing for spin exchange sampling. For $J_2 = 0$ one recovers the anti-ferromagnetic Heisenberg model and the system is frustration free with long-range Neel order \cite{sandvik1997finite, buonaura1998numerical}. In 1D the model is exactly solvable at the Heisenberg point ($J_2 = 0$) as well as for $J_2 = 0.5$, where it becomes the Majumdar-Gosh (MG) model \cite{majumdar1970antiferromagnetic} and undergoes a phase transition to a frustrated phase. A second phase transition from a gapless spin-fluid regime to a gapped dimer regime is found within the ferromagnetic regime at $J_2 \approx 0.24$ \cite{farnell1994coupled,bishop1998phase}. The model is known to be non-stoquastic for $J_2\neq 0$, and poses significant challenges for VMC \cite{park2020neural}.

\subsubsection{Chemistry Hamiltonians} \label{sec:chemcistry-hamiltonians}
Finally, we consider molecular Hamiltonians in second-quantized form 
\begin{align}
    \mathcal{H} = \sum_{i,j} Q_{ij} \cop_i \aop_j + \sum_{i,j,k,l} R_{ijkl} \cop_i \cop_k \aop_m \aop_j, \label{eq:chem-ham}
\end{align}
where $\cop_i$ and $\aop_j$ are fermionic creation and annihilation operators and $Q_{ij}$ and $R_{ijkl}$ are one and two-body integrals. Interacting fermionic Hamiltonians can be mapped to  spin systems by using e.g. the Jordan-Wigner (JW) or the Bravyi-Kitaev (BK) transformation \cite{bravyi2002fermionic,tranter2015b}. In the context of chemistry, NQSs using the JW transformation have been reported to yield more accurate VMC energies than those using the BK transformation \cite{choo2019fermionic}. Thus, we will concentrate on the JW transformation, which is formally given as
\begin{align}
    \aop_j &\rightarrow \bigg (\prod_{i=0}^{j-1} Z_i\bigg) P_j^{-} \label{eq:jw-I}\\
    \cop_j &\rightarrow \bigg (\prod_{i=0}^{j-1} Z_i\bigg) P_j^{+}, \label{eq:jw-II}
\end{align}
where $2 P_j^{\pm} = X_j\pm i Y_j$. Applying the transformation allows for mapping a Hamiltonian of the form given in equation \eqref{eq:chem-ham} to an interacting spin Hamiltonian such that we can train RBM quantum states to represent their ground state.
\section{Numerical Results}
For our simulations we used the open source library \textit{NetKet} \cite{netket:2019} (version 2.0.1) where we added our LM implementation. Details about hyperparameters during training as well as further numerical details can be found in the appendix \ref{app:numerical-details}.
\subsection{Scaling Analysis}
\subsubsection{Training Epochs}
\begin{figure} [t]
    \centering
    \includegraphics[width=\linewidth]{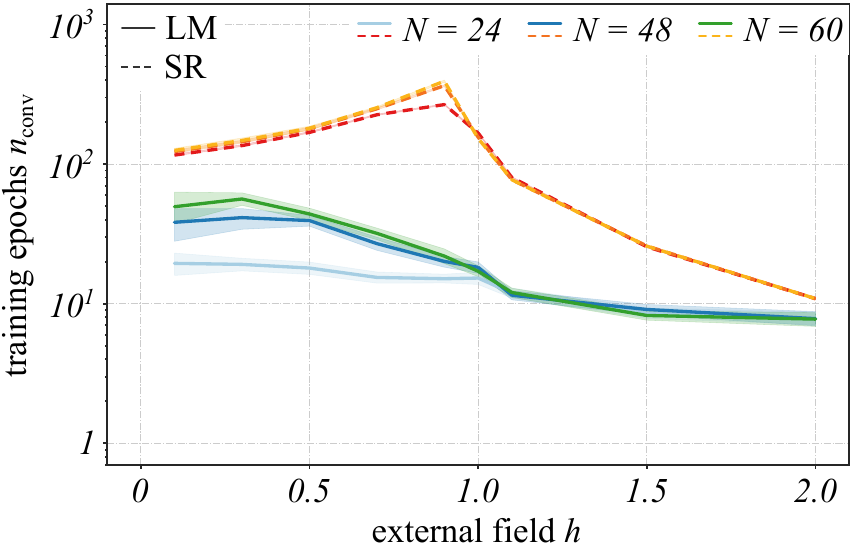}
    \caption{\textbf{Convergence epochs with LM and SR for the TFI model:} The number of epochs until convergence $\nis$ for the LM and the SR as a function of the external field $h$, where different colors encode different numbers of sites $N$ in the chains. The number of  iterations until convergence $\nis$ is determined as the VMC iteration $i$ for which the relative error $\erel(i) \leq 2\times 10^{-3}$ for the first time. For the point $h = 0$ we found that the runs either converged after very few iterations ($\nis \lesssim 3$) or did not converge at all. Thus, we removed $h = 0$ from our analysis.}
    \label{fig:convergence-iters}
\end{figure}

The learning curves for the TFI model in figure \ref{fig:vmc-curves-accuracy-lattice} suggest that training with the LM can require up to an order of magnitude fewer epochs than the SR to achieve comparable accuracies. This is illustrated more extensively in figure \ref{fig:convergence-iters} where the number of epochs for convergence is plotted for various values of $h$ and system sizes $N$.

We see that the LM convergences in a number of update steps which is typically an order of magnitude smaller than for SR. This is true of other second order methods, like the Newton-Raphson method. However, the LM is generally cheaper as second order derivatives do not need to be computed.

Moreover one observes a qualitatively different behavior of $\nis$ as a function of the external field $h$ for SR and LM when approaching the quantum critical point $h=1$. The number of epochs for convergence reaches a maximum near the critical point for SR, whereas the LM seems to be quite insensitive.  This might indicate that in addition to faster convergence properties the LM is less sensitive \wrt quantum criticality.

\subsubsection{Update Algorithm}
Looking at the bare number of training epochs, however, is only half the truth in neural network training. The final quantity of interest is the clock time required to find a proper approximation for the ground state wave function. Indeed the LM has a much larger per epoch cost, which can in certain circumstances cancel the convergence speedup. To approach this issue we define the time for a single epoch as the sum of the time required for sampling (algorithmic step (1)) and the time required for the calculation of the update step (algorithmic step (2) and (3)) 
\begin{align}
    \ti = \ts + \tu. \label{eq:tot-epoch-time}
\end{align}
For the symmetry encoding RBM the time scaling for an LM update step is dominated by the calculation of the local energy derivatives, which scales as $\mathcal{O}\big(\alpha \times N^2\big)$. For the SR, however, the time per epoch scales with the iterative solver used for the solution to \eq \eqref{eq:sr-update}, which erases the need for both, building the Fisher matrix and the inverse. It scales as $\mathcal{O}\big(N_\mathrm{cg} \times \alpha \times N\big)$, where $N_\mathrm{cg}$ is the number of iterations until convergence for the conjugate gradient solver. Often $N_\mathrm{cg}$ is much smaller than $N$ such that the calculation of the update step within the SR scales better than within the LM. For a  more detailed discussion of the computational complexities which are involved in the NQS training see appendix  \ref{app:computational-complexities}.
\begin{figure} [b]
    \centering
    \includegraphics[width=\linewidth]{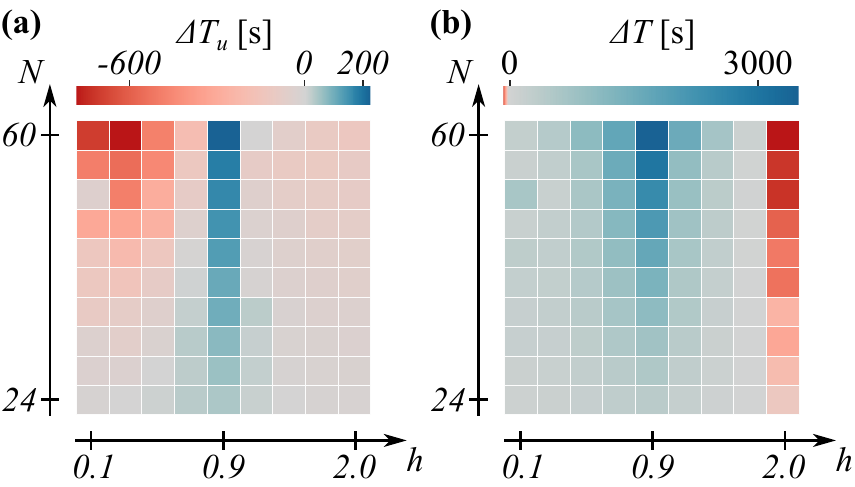}
    \caption{\textbf{(a) Runtime Phase Diagram for the Total Update Time in the TFI model: }Each tile is colored according to the difference in the total update time $\Delta \Tu = \Tu^\mathrm{sr} - \Tu^\mathrm{lm}$. The total update time describes the clock time until convergence, which is required for the calculation of the update step within the SR and the LM, respectively (see \eq \eqref{eq:total-update-time}). Essentially, this difference describes the training time difference between SR and LM when sampling is not taken into account. Blue tiles ($\Delta \Tu > 0$) correspond to regions where the total update time for the LM until convergence is smaller and vice versa. \textbf{(b) Runtime Phase Diagram for the Total Training Time in the TFI model: }Here the same situation as in (a) with the time for sampling taken into account is shown. The difference in total training time is defined as $\Delta \Tt = \Tt^\mathrm{sr} - \Tt^\mathrm{lm}$, where the total training time is the clock time required for the training of the NQS (see \eq \ref{eq:total-training-time}). Thus, it allows to identify the more efficient training algorithm. Again blue tiles are regions in which the LM is more efficient.}
    \label{fig:phase-diagrams-of-usage-tfi}
\end{figure}
Recalling that the LM needs approximately an order of magnitude fewer training iterations $\nis$ than the SR, we approach the issue of how the different scaling in computational time per epoch finds expression in the total update time $\Tu$. It is defined as 
\begin{align}
    \Tu = \sum_{i=1}^{\nconv} \tu^{(i)}, \label{eq:total-update-time}
\end{align}
where $\tu^{(i)}$ is the time for the update calculation at training epoch $i$. Thus the total update time $\Tu$ scales linearly in the number of  epochs. 
Figure \ref{fig:phase-diagrams-of-usage-tfi}.a shows what we call the \textit{runtime phase diagram} for the total update time. It serves the purpose of globally identifying regions in which either the LM or the SR requires less total update time $\Tu$. Therefore, each tile (corresponding to a combination of external field $h$ and system size $N$) is colored according to the difference $\Delta \Tu = \Tu^{\mathrm{sr}} - \Tu^{\mathrm{lm}}$ such that $\Delta \Tu > 0$ (blue) corresponds to regions in which the LM is faster \wrt the total update time. We find that the total update time for SR is shorter for most system sizes $N$ and external field values $h$ where one observes the difference to increase with increasing system size indicating asymptotic advantage. Only for an external field value of $h = 0.9$ do we see superior LM performance. Note that that each row in the runtime phase diagram corresponds to a plot like the one in figure \ref{fig:convergence-iters} where the number of training epochs is replaced by the total update time. Thus, when looking only at the time that is required for the update algorithm itself, the SR is found to be widely superior in the asymptotic limit.

\subsubsection{Total Training}
So far, the time for sampling has been disregarded in the scaling analysis in order to make a clear statement about the scaling relation of LM and SR as stand alone components. Clearly, the quantity of ultimate interest, however, is the total time that is required for the training of NQSs
\begin{align}
    \Tt = \sum_{i=1}^{\nconv} \ts^{(i)} + \tu^{(i)}. \label{eq:total-training-time}
\end{align}
Our simulations show clear evidence that, when taking sampling into account, the more efficient algorithm shifts from the SR to the LM (compare figure \ref{fig:phase-diagrams-of-usage-tfi}.a and \ref{fig:phase-diagrams-of-usage-tfi}.b). For the TFI model the LM converges to the ground state in less training time than the state-of-the art SR algorithm across most of the $N-h$-space (figure \ref{fig:phase-diagrams-of-usage-tfi}.b). Moreover, we find the training time difference to increase when going to larger system sizes suggesting an asymptotic training time advantage for the LM.
\subsection{Efficiency Regimes}
\begin{figure}
    \centering
    \includegraphics[width=\linewidth]{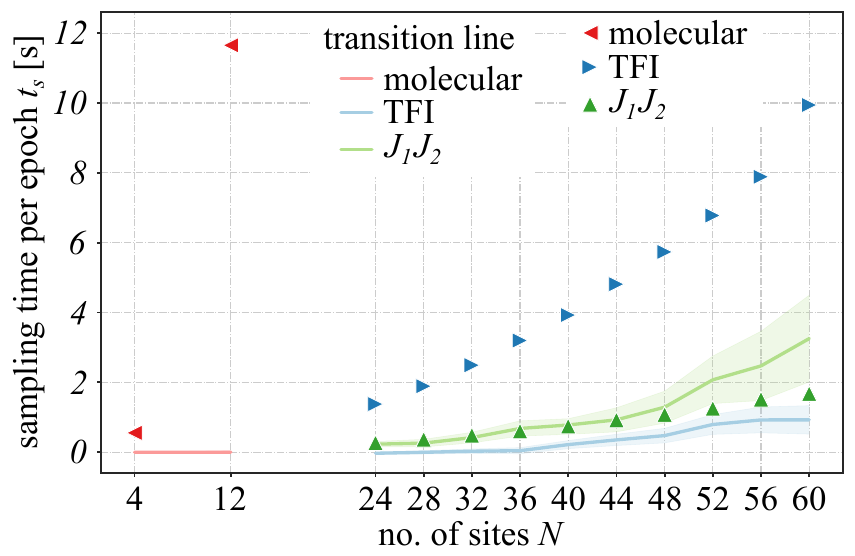}
    \caption{\textbf{Efficiency Regimes: }The sampling time per epoch $\ts$ for the J1J2 model (green triangles), the TFI model (blue triangles) and molecular Hamiltonians (red triangles) in terms of system size. For the molecular Hamiltonians $N = 4$ corresponds to $\mathrm{H}_2$ and $N = 12$ to $\mathrm{LiH}$, which have been transformed into spin basis by using the Jordan-Wigner transformation (see \eq \eqref{eq:jw-I} and \eqref{eq:jw-II}). 
    The solid lines describe the transition point $\ttrans$ of the epoch sample time, which is a function of the training epochs $\nconv$ and the update time per epoch $\tu$ for LM and SR, respectively. It describes the transition point for $\ts$, such that the LM is more efficient if $\ts > \ttrans$. If $\ts < \ttrans$ the SR is more efficient.} 
    \label{fig:efficiency-regimes}
\end{figure}

The scaling analysis from above suggests that the sampling cost is the determining factor separating the efficiency of the SR and LM methods. 
This is clearly seen in figure \ref{fig:efficiency-regimes}, when comparing the sampling time $\ts$ for the J1J2 model, TFI model and chemistry Hamiltonians.

As the ground state of the J1J2 model lies in the $M_z^0$ symmetry sector, MCMC exchange sampling can be used, which is computationally more efficient than single spin flip sampling,  used for the TFI model (green and blue triangles in figure \ref{fig:efficiency-regimes}). 
A vastly more expensive sampling cost arises in the context of molecular Hamiltonians (red triangles in figure \ref{fig:efficiency-regimes}), the reason for which is twofold: First, particle number and spin conservation has to be respected throughout the sampling. To accomplish that we use the \texttt{MetropolisHamiltonian} sampler from the package \textit{NetKet} \cite{netket:2019}. Second, the total number of samples required is of the order $\mathcal{O}(\Delta k \times 10^4)$, where $\Delta k = 10N$ is the downsampling interval. This is due to the sharply peaked nature of the absolute square of the wave function around the Hartree-Fock state \cite{choo2019fermionic}, which results in low acceptance probabilities during sampling. 
\begin{figure*}
    \centering
    \includegraphics[width=\linewidth]{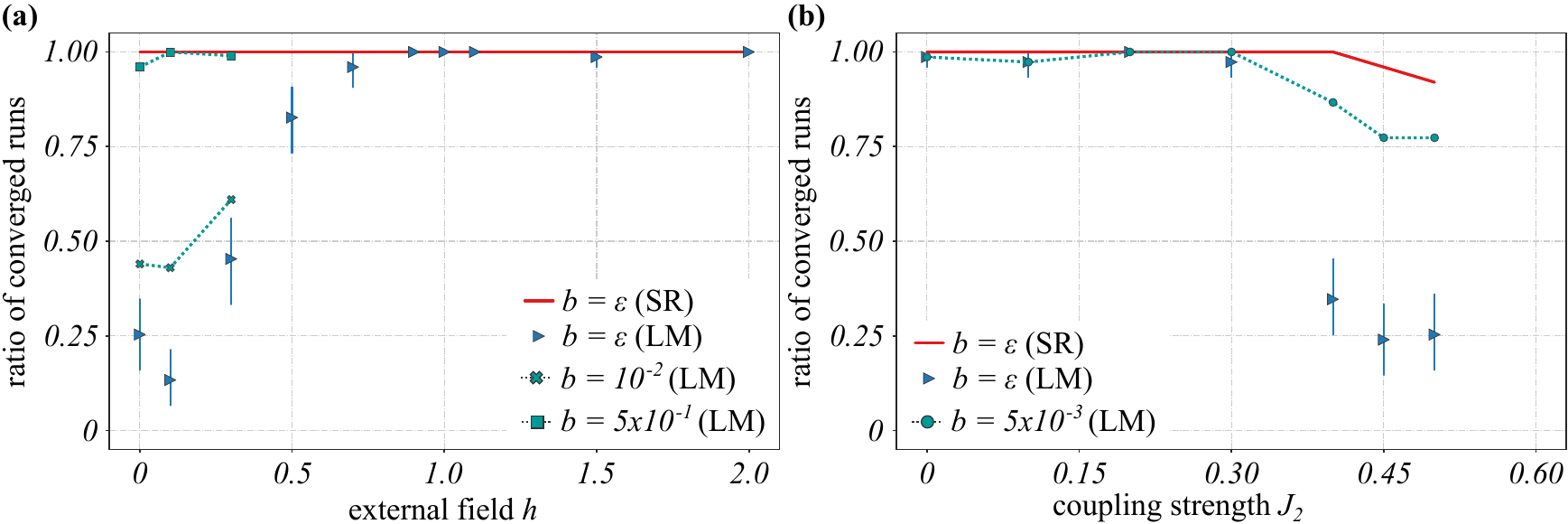}
    \caption{\textbf{Learning Reliability of Linear Method and Stochastic Reconfiguration: } Figure shows the learning reliability of LM and SR for the TFI \textbf{(a)} and the J1J2 model \textbf{(b)}. The learning reliability is defined as the probability for which an NQS training run converges to the ground state (see \eq \eqref{eq:conv-ratio}). Convergence is defined as $\erel \leq b$ and $b = \epsilon = 2\times10^{-3}$. Blue triangles correspond to the learning reliability of the LM, whereas red lines correspond to the learning reliability of the SR. The vertical bars are the $2\sigma$ confidence interval. Dashed turquoise lines are the probabilities of success when increasing the threshold $b$.}
    \label{fig:convergence-ratios}
\end{figure*}
Based on the number of training epochs $\nconv$ and the update time per epoch $\tu$ one can define a transition sampling time $\ttrans$ (solid lines in figure \ref{fig:efficiency-regimes}). It is defined such that for a model with sampling times per epoch $\ts < \ttrans$ the SR is more training efficient. For models in  which $\ts > \ttrans$ the LM is more efficient. Here it should be noted that we averaged the number of training epochs over all Hamiltonian parameters ($h$ for TFI and $J_2$ for J1J2 model) for each size $N$. Thus, the transition line can be understood as representing the model across the given Hamiltonian parameter space. The transition sampling time can be readily calculated by solving the equality 
\begin{align}
    \Tt^\mathrm{sr} - \Tt^\mathrm{lm} = 0,
\end{align}
which yields
\begin{align}
    \ttrans = \frac{\nconv^\mathrm{lm}\tu^\mathrm{lm}-\nconv^\mathrm{sr}\tu^\mathrm{sr}}{\nconv^\mathrm{sr} - \nconv^\mathrm{lm}}.
\end{align}

We interpret the gap between sampling time per epoch and the transition line $\ttrans$ as a measure for how much more efficient either of the training algorithms is for a model. Within our numerical simulations we found the SR to be the more efficient training algorithm for the sampling cheap J1J2 model, as the sampling time per epoch $\ts$ lies below the transition line (green triangles and green solid line in figure \ref{fig:efficiency-regimes}). When sampling becomes more expensive as is the case for the TFI model, we find the LM to be more training efficient (blue triangles above blue solid line). As the gap between $\ts$ and $\ttrans$ grows in the number of sites we expect the LM to be asymptotically more training efficient than the SR. This is indeed consistent with our findings from the runtime phase diagram in the section above (see also figure \ref{fig:phase-diagrams-of-usage-tfi}.b).

For molecular NQSs, our numerical results indicate a strong training time advantage for the LM, where $N = 4$ and $N = 12$ correspond to the qubit representations (see \eq \eqref{eq:jw-I} and \eqref{eq:jw-II}) of the $\mathrm{H}_2$ and the $\mathrm{LiH}$ molecule, respectively. Having the transition line close to zero furthermore indicates that the LM and the SR are nearly equally efficient when sampling is not taken into account. This is due to the fact, that the number of training epochs for molecular Hamiltonians with the LM is more than an order of magnitude smaller than with the SR. 

Finally, it is important to note that any improvements on the sampling strategies for specific models can tip the balance in favor of the SR method. 
\section{Linear Method for NQS Training}
\subsection{Learning Reliability}
In order to determine the learning reliability, we consider a VMC run to be converged if
\begin{align}
    \erel \leq b, \label{eq:conv-criteria}
\end{align}
meaning if the relative error is below some threshold $b$. If not stated otherwise we choose $b = \epsilon = 2 \times 10^{-3}$. The relative error is given as 
\begin{align}
    \erel = \frac{|E_0 - \evmc|}{|E_0|}, \label{eq;rel-error}
\end{align}
where $E_0$ is the exact ground state energy. 
The convergence ratio is calculated as the fraction
\begin{align}
    \convr = \frac{\Nvmc^*}{\Nvmc}, \label{eq:conv-ratio}
\end{align}
where $\Nvmc^*$ is the number of VMC runs that are labeled as converged according to \eq \eqref{eq:conv-criteria} and the total number of VMC runs is denoted as $\Nvmc$.
\begin{figure*}
    \centering
    \includegraphics[width=\linewidth]{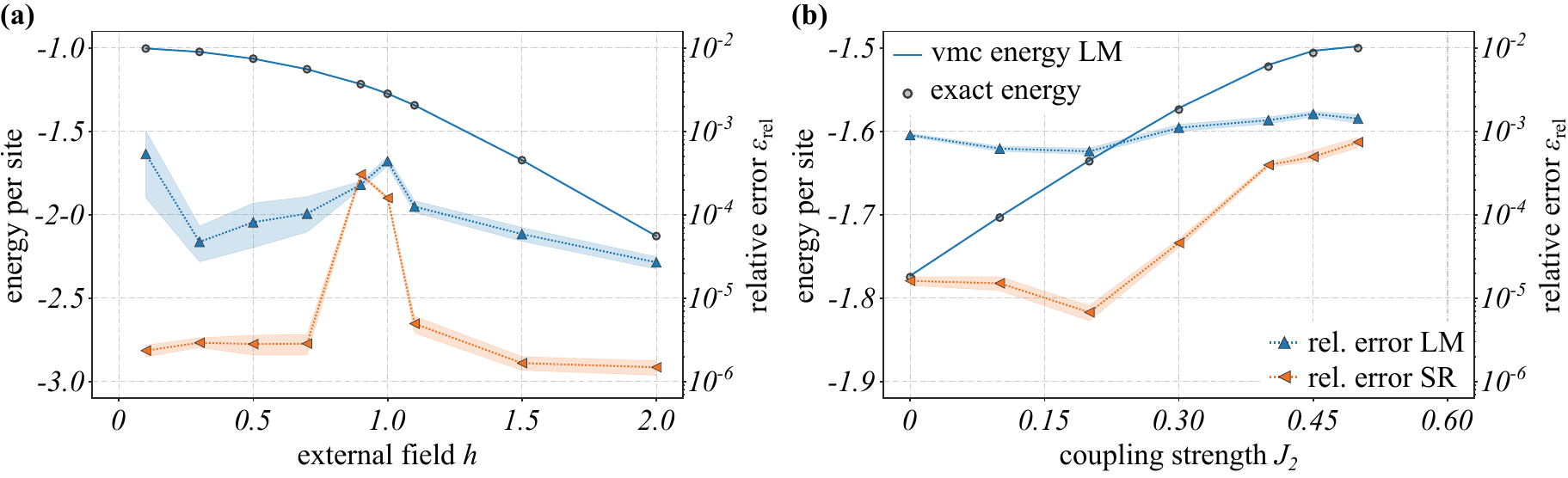}
    \caption{\textbf{Relative Errors for Linear Method and Stochastic Reconfiguration: } Figure shows the relative error and energy per site obtained with the LM (blue dotted lines) and the SR (orange dotted lines) for the TFI \textbf{(a)} and the J1J2 \textbf{(b)} model. The solid blue lines correspond to the energy per site predicted by NQS simulations with the LM. Black circles are reference energies. Transparent envelopes describe the $2\sigma$ confidence interval. \textbf{(a) Transverse Field Ising Model: } Simulations were run for a chain with $N = 40$ sites and various external field values $h$. Reference energies were obtained by mapping the TFI Hamiltonian to free fermions \cite{pfeuty1970one}. \textbf{(b) J1J2 Model: } Simulations were run for a chain with $N = 40$ sites and different values of next nearest neighbor coupling $J_2$. Reference energies were obtained by density matrix renormalization group calculations using the library \textit{iTensor} \cite{itensor}.}
    \label{fig:accuracy}
\end{figure*}
We find high convergence reliability for $h \geq 0.3$ and most values of $J_2$, but also observe regions in the model parameter space where $\convr \leq 0.25$ (figure \ref{fig:convergence-ratios}). For the TFI chain $\convr$ is particularly low close to the classical point $h = 0$ and increases rapidly when moving away from it. Importantly the LM shows high reliability at and around the quantum critical point $h = 1$. For the J1J2 chain $\convr$ drops if one is deep in the gapped dimer regime at $J_2 = 0.45$. Despite similar convergence ratios for $h \leq 0.3$ and $J = 0.45$, the reasons for the low convergence ratios of TFI and J1J2 model, however, originate from different causes. VMC runs that fail in the gapped dimmer regime usually come close to the global energy minimum but have a relative error slightly larger than $b = \epsilon$ and consequently are labelled as not converged. Increasing the error threshold up to $b = 5\times 10^{-3}$ leads to much higher values of $\convr$ in the gapped dimer regime, indicated by the dashed line in figure \ref{fig:convergence-ratios}.b. On the contrary, the non-convergent runs in the TFI get mostly stuck in local minima of the energy landscape with VMC energies far away from the real ground state energy. For that reason repeated convergence is only achieved when the threshold for the relative error is increased to $b = 0.5$ (dashed line with square markers in figure \ref{fig:convergence-ratios}.a). More generally, within the J1J2 model we observe that slightly increasing the threshold also increases the value of $\convr$ to one for $J_2 \leq 0.3$. This is consistent with our observation of larger fluctuations in the energy within the J1J2 chain than in the TFI chain (for $h \geq 0.5$). Compared to the SR (red lines in figure \ref{fig:convergence-ratios}) we find that the LM is less reliable for $b = \epsilon$. In particular, the SR shows also high reliability in regions in which the LM suffers from convergence ratios $\convr \leq 0.25$.
\subsection{Accuracy} \label{sec:accuracy}
Our results in figure \ref{fig:accuracy} show that the achieved accuracies with the LM strongly depend on the Hamiltonian parameters. In the TFI model we observe constant accuracies with a relative error $\erel \sim 10^{-4}$. However, for $h = 0.1$ we find the LM to yield less accurate results $\erel > 10^{-3}$ which is in line with our observations that the learning reliability is small for $h$ close to the classical point (figure \ref{fig:convergence-ratios}). Thus, this regime can be considered to be particularly hard for the LM. For the J1J2 model, we find the results for the ground state energy to be mostly less accurate than the ones observed in the TFI chain by approximately one order of magnitude. For $J_2 \leq 0.3$ the relative error is constant and increases when being deep in the gapped dimer regime. As for the TFI model this is consistent with a small convergence reliability $\convr$ at $J_2 = 0.45$. Compared to the SR, one finds the LM to be several orders of magnitude less accurate at $h = 2$. However, close to the quantum critical point $h=1$ as well as in the gapped dimer regime of the J1J2 model, we find the accuracy for the SR and the LM to be of similar order. 

\section{Discussion}
In this paper, we explored the LM as a learning algorithm for complex NQSs. We presented numerical evidence that the LM can be used for the learning of complex valued NQSs, to our knowledge for the first time. At the same time, we found the LM to deliver accurate representations for the ground state of the TFI and the J1J2 model on a one dimensional chain as well as for molecular Hamiltonians in a spin basis. We  observed the LM to converge in an order of magnitude fewer epochs than the state-of-the-art SR algorithm. As a consequence we found the LM to be the more time efficient training algorithm, whenever MCMC sampling becomes expensive in the NQS training. The particular strength of the LM is most evident on the example of molecular Hamiltonians, where we found it to be clearly advantageous compared to the SR \wrt training time. Thus, we expect the LM to be a valuable tool for the optimization of NQSs whenever sampling is expensive.

On the flip side, we observed a significantly larger variance for optimization with the LM which resulted in less accurate and less reliable results compared to the SR. Moreover, the LM showed a fragile nature during training when the optimization lacked an appropriate sign structure to start with, such that we encoded it by using the Marshall sign rule. This is indeed consistent with prior observations made during the optimization of NQSs \cite{choo2019two, park2020neural}, but we found it to be more severe for the LM than for the SR. Consequently, a deeper investigation regarding better phase learning with the LM might be a topic for future research.

With its origin in the chemistry community and thus usually applied to problems in first quantization, it might be an interesting direction to investigate the LM in the context of recently introduced NQSs in first quantization basis \cite{stokes2020phases}. Also in the context of chemical systems, the LM has been used for the direct optimization of excited states \cite{zhao2016efficient}, such that one could test an LM based optimization for excited NQSs similar to approaches with the SR \cite{choo2018symmetries}.
\section*{ACKNOWLEDGEMENTS}
The authors would like to thank Chae-Yeun Park and David Wierichs for helpful discussions. We furthermore thank the Regional Computing Center of the University of Cologne (RRZK) for providing computing time on the DFG-funded (Funding number: INST 216/512/1FUGG) High Performance Computing (HPC) system CHEOPS as well as support. The work presented here was completed while both authors were at the University of Cologne. 
\bibliography{Bibliography}
\newpage
\onecolumngrid
\appendix
\section{SR as Imaginary Time Evolution} \label{app:sr-as-imaginary-time-evolution}
Up to first order a single imaginary time evolution step is given as 
\begin{align}
    \ket{\wfi} = \left(1-\epsilon \mathcal{H}\right) \ket{\wfd{0}},
\end{align}
starting from the current wave function $\ket{\wfd{0}}$. Generally one can not assume that for any finite $\epsilon$ the wave function $\ket{\wfi}$ can be represented by the ansatz wave function. However, for $\epsilon$ chosen sufficiently small the wave functions $\ket{\psi_\epsilon}$ and $\ket{\wf}$ are close to each other. Then it is reasonable to assume that an update step $\dt$ exists such that the updated wave function $\ket{\wfe}$ is an approximate representation of $\ket{\psi_{\epsilon}}$. Thus, we write
\begin{align}
    (1-\epsilon \mathcal{H}) \ket{\wfd{0}} = e_0 \ket{\wfd{0}} + \sum_{k=1}^{\Nv} e_k \ket{\wfd{k}} + \ket{\psi_\perp}, \label{eq:app-cy-sr}
\end{align}
where $\{e_k\}$ are coefficients, $\ket{\wfd{k}} = \D{k} \ket{\wfd{0}}$ and $\ket{\psi_\perp}$ is a state of the orthogonal subspace \cite{park2020geometry}. Multiplying $\bra{\wfd{0}}$ and $\bra{\wfd{k}}$ from the left with equation \eqref{eq:app-cy-sr} gives
\begin{align}
     1 - \epsilon \braket{\mathcal{H}} &= e_0 + \sum_{k=1}^{\Nv} e_k \braket{\D{k}}\\
      \braket{\Dd{k}} - \epsilon \braket{\Dd{k}\mathcal{H} } &= e_0 \braket{\Dd{k}} + \sum_{k'=1}^{\Nv} e_{k'} \braket{\Dd{k} \D{k'}},
\end{align}
which is a set of linear equations for the coefficients. Solving for $e_0$ yields
\begin{align}
    \sum_{k'} S_{kk'} e_k = - \epsilon f_k, \label{eq:app-sr-equation}
\end{align}
with the $\bm{S}$ matrix
\begin{align}
    S_{kk'} = \braket{\Dd{k} \D{k'}} - \braket{\Dd{k}} \braket{\D{k'}}, \label{eq:app-s-matrix-sr}
\end{align}
and the force vector $\bm{f}$
\begin{align}
    f_k = \braket{\Dd{k} \mathcal{H}} - \braket{\Dd{k}}\braket{\mathcal{H}}. \label{eq:app-force-vector-sr}
\end{align}
Finally, we identify the coefficients $\{e_k\}$ as the update step $\dt$ such that the updates in the SR can be formally written as
\begin{align}
    \bm{\lambda}' \leftarrow \bm{\lambda} - \epsilon \bm{S}^{-1} \bm{f},
\end{align}
recovering equation \eqref{eq:sr-update} from the main text.
\section{Wave Function Phase Optimization} \label{app:wave-function-phase-optimization}
Let us start by considering the normalized wave function
\begin{align}
    \ket{\nwfdt{0}} = \frac{\ket{\wfd{0}}}{||\wfd{0}||}\label{eq:app-normalized-wf},
\end{align}
where we assume that $\ket{\wfd{0}}$ is holomorphic. Using Wirtinger derivatives $\pd$ and $\pdd$ the first order expansion of $\ket{\nwfdt{0}}$ can be written as (see below for full details)
\begin{align}
    \ket{\nwfet} = \ket{\nwfd{0}} + \sum_{k=1}^{\Nv} \ket{\nwfdt{k}} + O(|\dt|^2), \label{eq:app-wf-expansion-not-orthogonal}
\end{align}
where the change with respect to the $k$-th variational parameter is given as 
\begin{align}
    \ket{\nwfdt{k}} &= \frac{\mathcal{D}_k \ket{\wfd{0}}}{||\wfd{0}||} \dtk - \frac{\ket{\wfd{0}}}{||\wfd{0}||} \, \mathrm{Re}\left\{ \frac{\braket{\wfd{0}|\mathcal{D}_k|\wfd{0}}}{||\wfd{0}||^2} \dtk \right\}\nonumber \\
    &= \big[ \D{k} \dtk - \mathrm{Re}\left\{\braket{\D{k} } \dtk\right\} \big]\ket{\nwfdt{0}}\label{eq:app-derivative-normalized-wf}.
\end{align}
In equation \eqref{eq:app-derivative-normalized-wf}, $\braket{\D{k}} = \braket{\nwfdt{0}|\D{k}|\nwfdt{0}}$ denotes the expectation value of the log-derivatives (see \eq \ref{eq:log-derivatives} main text). Consequently, in the case of real parameters the current wave function and its derivatives build a self-plus tangent space as the orthogonality appears quite naturally as tangent space of a sphere in real space. Here we assumed that real network parameters also result in real variational wave functions, which is indeed the case for NQSs but e.g. does not hold for variational quantum eigensolvers (VQEs). On the contrary, complex parameters lead to a purely imaginary overlap between the current wave function and its derivatives 
\begin{align}
    \braket{\nwfdt{0}|\nwfdt{k}} = i \, \mathrm{Im} \left\{\braket{\D{k}}\dtk\right\},
\end{align}
indicating that $\ket{\nwfdt{k}} \notin \Omega$ as also suggested in \cite{motta2015implementation}. Consequently, the first order expansion of the normalized wave function $\ket{\nwfet}$ does not lie in $\Omega^+$ for the complex case in general. We will now show that it is possible to construct a first order expansion $\ket{\nwfe}$, which lies in $\Omega^+$. Starting point is the linear expansion given in equation \eqref{eq:app-wf-expansion-not-orthogonal} 
\begin{align}
     \ket{\overline{\psi}_{\bm{\lambda} + \dtk}} &= \bigg ( 1 + \dtk \mathcal{D}_k - \mathrm{Re} \big\{ \braket{\D{k}} \dtk \big\}  \bigg) \ket{\nwfdt{0}} + \Ord \nonumber\\
    &= \bigg( 1 + i \, \mathrm{Im}\big\{ \braket{\D{k}}\dtk \big\} \bigg) \ket{\nwfdt{0}} + \dtk \ket{\nwf{k}} + \Ord, \label{eq:app-lm-normalized-wf-I}
\end{align}
where we have introduced
\begin{align}
   \ket{\nwfd{k}} = \big[ \D{k} - \braket{\D{k}} \big]\ket{\nwfdt{0}}, \label{eq:app-tangent-space-derivative}
\end{align}
with $\ket{\nwfd{0}} \equiv \ket{\nwfdt{0}}$. It can easily be verified that $\ket{\nwfd{k}} \in \Omega$, by multiplying $\bra{\nwfdt{0}}$ from the right to \eqref{eq:app-tangent-space-derivative} and realizing that the overlap is indeed zero. 
Adding a term quadratic in $\delta \lambda_k$ to equation \eqref{eq:app-lm-normalized-wf-I}, such that the pre-factors become the first order Taylor expansion of the exponential function, allows to write \cite{becca2017quantum}
\begin{align}
     \ket{\overline{\psi}_{\bm{\lambda} + \dtk}} &= \bigg( 1 + i \, \mathrm{Im}\big\{ \braket{\D{k}} \dtk \big\} \bigg) \ket{\nwfd{0}} + \bigg( 1 + i \, \mathrm{Im}\big\{ \braket{\D{k}}\dtk \big\} \bigg) \dtk \ket{\nwfd{k}} + \Ord \nonumber\\
     &= \exp \big[ i \phi_k \big] \bigg( \ket{\nwf{0}} + \dtk \ket{\nwfd{k}} \bigg) + \Ord, \label{eq:app-expansion-nwf}
\end{align}
with $\phi_k = \mathrm{Im}\big\{ \braket{\D{k}} \dtk \big\}$. The general expression when several parameters are changed can easily be deduced from equation \eqref{eq:app-expansion-nwf} and is given by
\begin{align}
    \ket{\nwfe} = \exp \left[ i \Phi \right] \left( \ket{\nwfd{0}} + \sum_{k=1}^{\Nv} \dtk \ket{\nwfd{k}} \right) + O\big(|\dt|^2\big), \label{eq:app-expansion-wf-several-params}
\end{align}
where $\Phi = \sum_k \phi_k$, is a global phase factor. 

Choosing $\Phi$ such that the pre-factor becomes one (as it will not affect the following derivations), the first order expansion of the normalized wave function can be written as
\begin{align}
    \ket{\nwfe} = \sum_{n=0}^{\Nv}\ket{\nwfd{n}} \label{eq:app-lin-exp-wf}
\end{align}
where $\ket{\nwfd{n}} \in \Omega$ for $n > 0$ by construction and consequently $\ket{\nwfe} \in \Omega^+$. It should be noted, that the construction of $\ket{\nwfe} \in \Omega^+$ corresponds to the heuristic re-scaling procedure outlined in \cite{umrigar2007alleviation} with $\xi = 1$, which allows for the update of non-linear parameters. This underlines the importance of the derivatives to lie in $\Omega$, which thus is considered to be crucial for the stability of the LM \cite{feldt2020excited}.
\subsection{Linear Wave Function Expansion} \label{app:linear-wave-function-expansion}
Let us start by assuming a holomorphic ansatz wave function $\ket{\wf}$ such that the following relations hold:
\begin{align}
   \ket{\pd\wf} &= \big(\bra{\pdd\wf}\big)^\dagger, \label{eq:hol-def-I}\\
   \bra{\pd\wf} &= \big(\ket{\pdd\wf}\big)^\dagger = 0, \label{eq:hol-def-II}
\end{align}
where $\pd$ and $\pdd$ denote the Wirtinger derivatives \cite{fischer2005precoding}. Also we consider the explicitly normalized wave function $\ket{\nwfdt{0}}$ for the current parameter configuration as
\begin{align}
    \ket{\nwfdt{0}} = \frac{\ket{\wfd{0}}}{\sqrt{\braket{\wfd{0}|\wfd{0}}}}.
\end{align}
Following Wirtinger calculus the first order expansion of the normalized wave function can be written as \cite{hunger2007introduction}
\begin{align}
    \ket{\nwfet} = \ket{\nwfdt{0}} + \pd \ket{\nwfdt{0}} \dtk + \pdd \ket{\nwfdt{0}} \dtk^*,
\end{align}
where Einstein sum convention is used for simplicity. It should be noted that $\ket{\nwfdt{0}}$ is generally not holomorphic due to the normalization. Explicitly building the derivatives with respect to $\pd$ and $\pdd$ then results in
\begin{align}
    \ket{\nwfet} &= \ket{\nwfdt{0}} + \frac{\ket{\pd\wfd{0}} \braket{\wfd{0}|\wfd{0}}^{\sfrac{1}{2}} - \frac{1}{2} \frac{1}{\braket{\wfd{0}|\wfd{0}}^{\sfrac{1}{2}}}\ket{\wfd{0}}  \big(\pd \braket{\wfd{0}|\wfd{0}}\big)}{\braket{\wfd{0}|\wfd{0}}}\dtk \nonumber\\
    &\hspace{2em} +\frac{\ket{\pdd\wfd{0}} \braket{\wfd{0}|\wfd{0}}^{\sfrac{1}{2}} - \frac{1}{2} \frac{1}{\braket{\wfd{0}|\wfd{0}}^{\sfrac{1}{2}}}\ket{\wfd{0}}  \big(\pdd \braket{\wfd{0}|\wfd{0}}\big)}{\braket{\wfd{0}|\wfd{0}}}\dtk^*. \label{eq:beast}
\end{align}
By using the relations \eqref{eq:hol-def-I} and \eqref{eq:hol-def-II} from above, equation \eqref{eq:beast} can be simplified to
\begin{align}
    \ket{\nwfet} &= \ket{\nwfdt{0}} + \frac{\ket{\pd \wfd{0}} }{\sqrt{\braket{\wfd{0}|\wfd{0}}}} \dtk \nonumber\\ 
    &\hspace{5em}- \frac{1}{2} \frac{\ket{\wfd{0}} }{\braket{\wfd{0}|\wfd{0}}^{\sfrac{3}{2}}} \bigg( \braket{\wfd{0}|\pd \wfd{0}} \dtk + \braket{\pdd \wfd{0}| \wfd{0}} \dtk^* \bigg)\\
    &= \ket{\nwfdt{0}} + \frac{\ket{\pd \wfd{0}} }{\sqrt{\braket{\wfd{0}|\wfd{0}}}}\dtk - \frac{\ket{\wfd{0}} }{\sqrt{\braket{\wfd{0}|\wfd{0}}}} \mathrm{Re}\bigg\{ \frac{\braket{\wfd{0}|\pd \wfd{0}}}{\braket{\wfd{0}|\wfd{0}}} \dtk \bigg\}.
\end{align}
Recalling the definition of the log-derivatives as $\D{k} \ket{\wfd{0}} = \ket{\pd \wfd{0}}$ one can rewrite the equation from above as
\begin{align}
    \ket{\nwfet} = \ket{\nwfdt{0}} + \bigg(\D{k} \dtk - \mathrm{Re}\big\{ \braket{\D{k}} \dtk \big\}\bigg) \ket{\nwfdt{0}}.
\end{align}
\section{Stochastic Estimation of Quantum Expectation Values} \label{app:stochastic-estimation-of-quantum-expectation-values}
\subsection{Stochastic Reconfiguration}
The quantum expectation values appearing in the entries for the $\bm{S}$ matrix and the force vector $\bm{f}$ can be estimated as \begin{align}
    S_{kk'} = \sexpb{\Dd{k} \D{k'}} - \sexpb{\Dd{k}} \sexpb{\D{k'}},
\end{align}
and
\begin{align}
    f_k = \sexpb{\Dd{k} \mathcal{\Hl}} - \sexpb{\Dd{k}}\sexpb{\mathcal{\Hl}},
\end{align}
which corresponds to just replacing the quantum expectation values $\braket{\cdot}$ in equations \eqref{eq:s-matrix-sr} and \eqref{eq:force-vector-sr} by their statistical counterparts $\sexpb{\cdot}$.
\subsection{Linear Method}
Clearly, no estimates are necessary for the first row and columns of the overlap matrix, where the stochastic estimator for the remaining part is given as
\begin{align}
    \overline{S}_{kk'} &= \sexpb{\Dd{k} \D{k'}} - \sexpb{\Dd{k}} \sexpb{\D{k'}},
\end{align}
where $k,k' > 0$. For the Hamilton matrix, the entries can be estimated
\begin{align}
    H_{00} &= \sexpb{\Hl},\\
    H_{k0} &= \sexpb{\Dd{k} \Hl} - \sexpb{\Dd{k}}\sexpb{\Hl},\\
    H_{0k'} &= \sexpb{\Hld{k'}} + \sexpb{\Hl \D{k'}} - \sexpb{\Hl} \sexpb{\D{k'}},\\
   H_{kk'} &= \sexpb{\Dd{k} \Hl \D{k'}}  - \sexpb{\Dd{k} \Hl} \sexpb{\D{k'}} - \sexpb{\Dd{k}} \sexpb{\Hl \D{k'}} \nonumber\\&+ \sexpb{\Dd{k}} \sexpb{\Hl} \sexpb{\D{k'}} + \sexpb{\Dd{k} \Hld{k'}} -  \sexpb{\Dd{k}} \sexpb{\Hld{k'}},
\end{align}
where $H_{00}$ and $H_{k0}$ are readily obtained from equations \eqref{eq:lm-h00} and \eqref{eq:lm-h0k}. A little more thought has to be given to the case when estimating the remaining entries of the Hamilton matrix, due to quantum expectation values of the form $\braket{\mathcal{H}\D{k}} \equiv \braket{\mathcal{H}_k}$. As it can be readily verified they are directly related to the local energy derivatives by $\Hld{k}(\bx) = \mathcal{H}_k(\bx) - \Hl(\bx)\D{k}(\bx)$, clarifying their appearance in the estimator equations above. At first glance, introducing the derivatives of the local energy seems like an overly complicated way for estimating $\bm{H}$, as the total number of terms increases compared to the estimator written in terms of quantum expectation. However, writing the estimators in terms of the local energy derivatives leads to statistical estimators, which are either a covariance or a tri-covariance, reducing statistical fluctuations and thus lowering the amount of total MC samples needed \cite{umrigar2005energy}.
\section{Computational Complexity} \label{app:computational-complexities}
In the following we will denote the number of sites as $N$, the network parameters as $\Nv$ and the number of MCMC samples as $\Ns$. The scaling for calculating the amplitude $\wfrbm(\bx)$ for given state $\bx$ is given as $\O{M}$ and the scaling for calculating the log-derivatives as $\O{\Nv}$.
\subsection*{Sampling}
Within the MH algorithm for each spin flip attempt the acceptance probability has to be calculated as 
\begin{align}
        A\big(\bx\rightarrow \bx'\big) = \mathrm{Min} \left(1, \left|\frac{\wfrbm(\bx')}{\wfrbm(\bx)}\right|^2\right),
\end{align}
and thus scales as $\O{M}$. Assuming $\O{N}$ spin flip attempts, a single MC sweep scales as $\O{N \times M}$. Consequently, generating in total $\Ns$ MC samples for the stochastic estimation of quantum expectation values scales as $\O{\Ns \times N \times M}$.
\subsection*{Local Energy}
The local energy for a given spin configuration $\bx$ can be written as
\begin{align}
    \Hl(\bx) = \sum_{\bx} \braket{\bx|\mathcal{H}|\bx'} \frac{\wfrbm(\bx')}{\wfrbm(\bx)}.\label{eq:ann-eloc}
\end{align}
The number of non-zero matrix elements of $\mathcal{H}$ in the basis $\bx$ grows as $\O{N}$ for lattice spin systems (note that this changes for the chemical Hamiltonians (\S\ref{sec:chemcistry-hamiltonians}) where $N$ has to be replaced by the number of Pauli strings after applying the JW transformation). The calculation of the wave function ratio scales as the calculation of $\wfrbm(\bx)$ as $\O{M}$. Consequently, calculating equation \eqref{eq:ann-eloc} for given state $\bx$ scales as $\mathcal{O}\big( N \times M \big)$. The calculation for $\sexpb{\Hl}$ scales then as $\mathcal{O}\big(\Ns \times N \times M\big)$.
\subsection*{Log-derivatives}
The scaling of the log-derivatives itself scales as $\O{\Nv}$ (by definition from above). Consequently, the stochastic estimation of all log-derivatives scales as $\O{\Ns \times \Nv}$. 
\subsection*{Local Energy Derivatives}
Recalling that the local energy derivatives are given as
\begin{align}
    \Hld{k}(\bx) = \mathcal{H}_k(\bx) - \Hl(\bx)\D{k}(\bx), \label{eq:ann-relation-hlocder-hk}
\end{align}
where $\mathcal{H}_k(\bx)$ is given as
\begin{align}
     	\mathcal{H}_k(\bx) = \sum_{\bx'} \braket{\bx|\mathcal{H}|\bx'} \mathcal{D}_k(\bx') \frac{\wfrbm({\bx'})}{\wfrbm(\bx)}.
\end{align}
Consequently, the calculation of all local energy derivatives for given $\bx$ scales as $\O{N \times \Nv \times M}$. The stochastic estimate scales then as $\O{\Ns \times N \times \Nv \times M}$.
\subsection*{Stochastic Reconfiguration}
Within the SR, one has to construct the quantum Fisher matrix $\bm{S}$ and the force vector $\bm{f}$ which have the dimensions $(\Nv\times\Nv)$ and $(\Nv \times 1)$, respectively. Consequently, their construction scales as $\O{\Nv^2}$ and $\O{\Nv}$. The update step is found as solution to a linear set of equations build by $\bm{S}$ and $\bm{f}$ (\S\ref{app:sr-as-imaginary-time-evolution}). Its solution can found by building the inverse $\bm{S}^{-1}$, which scales as $\O{\Nv^3}$.
\subsubsection*{Iterative Solver} \label{app:iterative-solver}
In order to speed things up, one can make use of an iterative algorithm proposed in \cite{neuscamman2012optimizing}, which erases the need for explicitly building the $\bm{S}$ matrix as well as calculating its inverse. Instead, the linear set of equations is solved by an iterative conjugate gradient method, which requires a computational cost of $\mathcal{O}\big( N_\mathrm{var}\big)$ per step, due to the product structure of $\bm{S}$ \cite{carleo2017solving}. Usually, the number of CG iterations is much smaller than the number of variational parameters such that using an iterative CG scheme greatly reduces the computational cost of the SR algorithm.
\subsection*{Linear Method}
The construction of the Hamilton $\bm{H}$ and overlap matrix $\overline{\bm{S}}$ scales as $\O{\Nv^2}$ thus is in line with the scaling for the construction of $\bm{S}$ in the SR. However, as the LM requires calculation of the local energy derivatives (see \eqref{eq:ann-relation-hlocder-hk}) the calculations of the matrix entries itself is computationally more expensive than in the SR. The update step is found as solution to the generalized eigenvalue problem of dimension $\Nv\times\Nv$, which generally scales as $\O{\Nv^3}$ (build the inverse on one side and solve the resulting regular eigenvalue equation with the Schur decomposition). The scaling for the eigenvalue problem can be reduced by using the Implicitly Restarted Arnoldi Method (IRAM).
\subsubsection*{Implicitly Restarted Arnoldi Method} \label{app:iram}
IRAM is particularly strong when one is interested in the outer eigenvalues/vectors of a system, which is the case in the LM (eigenvalue/vector with smallest real part). For an extensive introduction to theoretical and numerical details the reader is referred to \cite{sorensen1997implicitly} and \cite{lehoucq1998arpack}. As it is an iterative solver the actual gain in computational cost is hard to estimate and strongly depends on the hyper parameters and the structure of the matrices. To our knowledge finding appropriate upper bounds for the run time of IRAM is still an open question.
\section{Numerical Details} \label{app:numerical-details}
We want to stretch the fact that our results have been observed with NetKet version 2.0.1. In the meantime major changes \wrt the code architecture have been performed including a transition from C\texttt{++} to Python. Thus, the reported times are likely to differ even when using identical architectures and hyperparameters. The simulations were performed on the CHEOPS cluster at RRZK Cologne on Intel XEON X5550/X5560 processors which have a base operating frequency of $2.66\, \mathrm{GHz}$ and a maximum operating frequency of $3.06\, \mathrm{GHz}$.
\subsection{Network Architecture}
For simulating the TFI and the J1J2 model we used symmetry encoding RBM quantum states with hidden unit density of $\alpha = 2$. 

The chemistry Hamiltonian simulations were performed without the symmetry encoding RBM architecture and also with a hidden unit density of $\alpha = 2$.
\subsection{Training Hyperparameters}
The following table lists the training hyperparameters used for the data shown in the different figures.
\begin{center}
\begin{tabular}{ |c|c|c|c|c|c|c|c| } 
 \hline
 Figure & Model & Sampler & No. of samples $\Ns$ & Optimizer & Diag. shift $a_\mathrm{diag}$ & Learning rate $\eta$ & Tikhonov reg. $\kappa_0$ \\ 
 \hline
 \multirow{2}{*}{Fig.1/2/3/4} & \multirow{2}{*}{TFI} & \multirow{2}{*}{Local} & \multirow{2}{*}{$10^3$} & LM & 0.01 & --- & 0.5\\
 & & & & SR & 0.01 & 0.01 & --- \\
 \hline
 \multirow{8}{*}{Fig.5} & \multirow{2}{*}{TFI} & \multirow{2}{*}{Local}& \multirow{2}{*}{$10^3$} & LM & 0.01 & --- & 0.5 \\ 
 & & & & SR & 0.01 & 0.01 & --- \\
 & \multirow{2}{*}{J1J2} & \multirow{2}{*}{Exchange} & \multirow{2}{*}{$10^3$} & LM & 0.01 & --- & 0.5 \\ 
 & & & & SR & 0.01 & 0.01 & --- \\
 & \multirow{2}{*}{$\mathrm{H}_2$} & \multirow{2}{*}{Hamiltonian} & \multirow{2}{*}{$10^4$} & LM & 0.01 & --- & 0.1 \\ 
 & & & & SR & 0.01 & 0.05 & --- \\
 & \multirow{2}{*}{$\mathrm{LiH}$} & \multirow{2}{*}{Hamiltonian} & \multirow{2}{*}{$10^4$} & LM & 0.01 & --- & 0.001 \\ 
 & & & & SR & 0.01 & 0.05 & --- \\
 \hline
 \multirow{4}{*}{Fig.7/8} & \multirow{2}{*}{TFI} & \multirow{2}{*}{Local}& \multirow{2}{*}{$10^3$} & LM & 0.01 & --- & 0.5 \\ 
 & & & & SR & 0.01 & 0.01 & --- \\
 & \multirow{2}{*}{J1J2} & \multirow{2}{*}{Exchange} & \multirow{2}{*}{$10^3$} & LM & 0.01 & --- & 0.5 \\ 
 & & & & SR & 0.01 & 0.01 & --- \\
 \hline
\end{tabular}
\end{center}
\end{document}